\documentclass[doublecol]{epl2} 

\title{Mott transition in Cr-doped V$_2$O$_3$ studied by ultrafast reflectivity: electron correlation effects on the transient response}
\shorttitle{Mott transition in Cr-doped V$_2$O$_3$ studied by ultrafast reflectivity} 

\author{B. Mansart\inst{1} \and D. Boschetto\inst{2} \and S. Sauvage\inst{3} \and A. Rousse\inst{2} \and M. Marsi\inst{1}}
\shortauthor{B. Mansart \etal}

\institute{                    
  \inst{1} Laboratoire de Physique des Solides, CNRS-UMR 8502, Universit\'{e} Paris-Sud - F-91405 Orsay, France\\
  \inst{2} Laboratoire d'Optique Appliqu\'{e}e, ENSTA, CNRS, Ecole Polytechnique - 91761 Palaiseau, France\\
  \inst{3} Institut d'Electronique Fondamentale, CNRS-UMR 8622, Universit\'{e} Paris-Sud - F-91405 Orsay, France
}

\abstract{
The ultrafast response of the prototype Mott-Hubbard system (V$_{1-x}$Cr$_x$)$_2$O$_3$ was systematically studied with fs pump-probe reflectivity, allowing us to clearly identify the effects of the metal-insulator transition on the transient response. The isostructural nature of the phase transition in this material made it possible to follow across the phase diagram the behaviour of the detected coherent acoustic wave, whose average value and lifetime depend on the thermodynamic phase and on the correlated electron density of states. It is also shown how coherent lattice oscillations can play an important role in some changes affecting  the ultrafast electronic peak relaxation at the phase transition, changes which should not be mistakenly attributed to genuine electronic effects. These results clearly show that a thorough understanding of the ultrafast response of the material over several tenths of ps is necessary to correctly interpret its sub-ps excitation and relaxation regime, and appear to be of general interest also for other strongly correlated materials.  
}

\begin{document}

\maketitle
\section{Introduction}
\label{intro}
The physics of strongly correlated materials is witnessing the advent of a rapidly increasing number of ultrafast spectroscopy studies using various experimental setups. Ultrafast methods, based on the use of fs lasers in a pump-probe configuration, provide an original approach to the study of these materials, exploring in real time their excited states and their relaxation dynamics. This can provide important information on the basic physical parameters characterizing these systems and on the transitions in their thermodynamic phase diagram, frequently determined by complicated interplays among structural, electronic and magnetic effects~\cite{Imada1998,Dagotto2005}. Among these various spectroscopies, pump-probe optical reflectivity is probably the most direct approach to ultrafast science and is playing a major role in the study of strongly correlated systems ~\cite{Rini2007, 
Iwai2003, Averitt2002}. 

Extracting the relevant physical information from pump-probe reflectivity measurements can nevertheless be quite challenging: it usually involves identifying the aforementioned contributions to the overall signal, to be able to correctly measure important parameters like lifetime of excited states, electron-phonon coupling, lattice thermalization, and many others. It is consequently of great importance to understand in detail the ultrafast response of simple archetypical systems, in order to clarify the typical spectroscopic signatures of each of these individual effects present in correlated materials.   

This work focuses on the effects contributing to the ultrafast response of a prototype Mott-Hubbard compound, presenting a metal-insulator transition in its most essential form, i.e. without change in structure or in long range magnetic order, but only caused by the strong Coulomb repulsion between valence electrons and only resulting in a gap opening at the Fermi energy. In particular, we present the results of comprehensive series of pump-probe reflectivity measurements on vanadium sesquioxide (V$_{1-x}$Cr$_x$)$_2$O$_3$ across the Mott transition. In this system, the metal-Mott insulator transition can be obtained by applying an external pressure or changing the doping level $x$~\cite{Mott1990,McWhan1969}: the undoped compound, V$_2$O$_3$, is a strongly correlated metal at room temperature, and by introducing Cr-doping the system can cross the Mott transition and become insulating. This first-order transition does not involve any symmetry breaking, and the material maintains the same corundum structure. Due to these properties, numerous experimental and theoretical studies have been carried out on V$_2$O$_3$ starting from the late sixties~\cite{McWhan1970,Dernier1970}, and this material is often used to test the validity of the different models for metal-insulator transitions; also, several advances in experimental methods such as X-ray absorption~\cite{Park2000, Rodolakis2010}, transport~\cite{Limelette2003}, bulk-sensitive photoemission~\cite{Mo2003, Mo2006, Rodolakis2009} or optical spectroscopy~\cite{Baldassarre2008}, especially when combined with microscopic methods \cite{Lupi2010, Gunther1997}, have recently revived the interest on its Mott transition.

As already mentioned, time resolved spectroscopies offer appealing perspectives for the investigation of strongly correlated systems, since they give access to the dynamical excitation and relaxation of electrons close to the Fermi level. The possibility of exciting coherent optical or acoustic phonons and of observing them in real-time is also of great interest, since they are tightly related to the electronic and structural properties of the material. All these possibilities have been already extensively applied to study a prototype material for metal-insulator transitions accompanied by structural transitions, the sibling compound VO$_2$ \cite{Cavalleri2004, Kim2006}. For vanadium sesquioxide, the first time-resolved reflectivity measurement was performed on the undoped compound \cite{Misochko1998}, thus not exploring the Mott insulating part of the phase diagram; these results were confirmed in a more recent study on complex oxides, where the effects of crytallographic orientation with respect to the light polarization were discussed \cite{Mansart2010}.  
In this Letter, we clarify the effects on the ultrafast response of the most 
notable property of (V$_{1-x}$Cr$_x$)$_2$O$_3$, namely the isostructural 
metal-insulator transition between its Paramagnetic Metallic ($x=0$, PM) and Paramagnetic Insulating ($x=0.028$, PI) phases: the fact that the crystal structure is maintained is critically important in our experiment, because if a break in symmetry occurred, its disruptive effects on the coherent phonons \cite{Kim2006} would not allow an unambiguous comparative analysis. Moreover, our study covers the range from the femtosecond to the picosecond time scale, and is performed in a very large excitation density range, from 0.09 to 36.6 mJ/cm$^2$. The systematic nature of our experiments vs. various parameters (thermodynamic phase, initial temperature, laser fluence and wavelength) and the relative simplicity of the V$_2$O$_3$ phase diagram, allowed us to disentangle genuine phase difference from other effects. We found that it is particularly important to know the behaviour of the transient reflectivity over a wide time range (40 ps) to correctly interpret the response of the sub-ps window in its various components. These results, obtained on a prototype  isostructural metal-insulator transition, clarify the role of the correlated electron density of states in some features of the ultrafast response, and provide solid ground for the interpretation of other, more complex phase transitions in other materials. 

\section{Experimental}
\label{exp}
Time-resolved reflectivity measurements have been carried out on single crystals of (V$_{1-x}$Cr$_x$)$_2$O$_3$, $x=0, 0.028$. High quality single crystals from Purdue University were carefully oriented using Laue diffraction and mechanically polished to obtain mirror-like surfaces. We used two different setups, both of them based on a mode-locked Ti:Sapphire laser system. The first one, described in~\cite{Mansart2010, Boschetto2008}, uses a 1 kHz repetition rate system, providing pulses of $\approx$ 40 fs at 800 nm wavelength and high pump fluences ranging from 6.9 to 36.6 mJ/cm$^{2}$; the probe fluence was fixed at 0.95 mJ/cm$^{2}$ - well below the pump fluence - to avoid any perturbation of the system induced by the probing process. For this setup, the angle of incidence of the pump beam was almost normal with respect to the sample surface, while the probe arrived approximately at 15$^\circ$. The spot diameters (Full Width at Half Maximum) on the sample surface were 100 $\mu$m for the pump and 25 $\mu$m for the probe. The second setup was based on an Optical Parametric Oscillator, delivering pulses of $\approx$ 150 fs at $\approx$ 1550 nm. For this setup, the incidence angle of both pump and probe beams was normal to the sample surface; the excitation densities ranged from 0.09 to 0.35 mJ/cm$^{2}$ and probe fluence was fixed at 3.2 $\mu$J/cm$^2$. Pump and probe diameter (FWHM) on the sample surface were 2 $\mu$m. Both setups used optical choppers and orthogonal polarisations between pump and probe in order to maximize the signal-to-noise ratio.

\section{Results and discussion}
\label{disc}

\begin{figure}[htb] 
\begin{center}
\includegraphics[width=1\linewidth,clip=true]{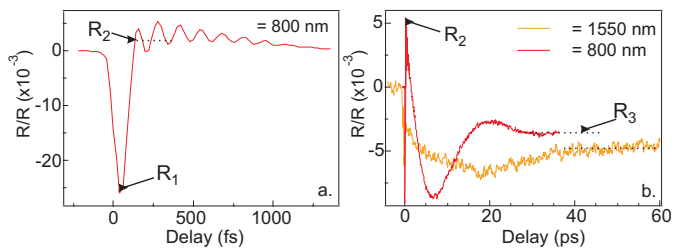} 
\caption{(Colour online) Time resolved reflectivity in two different time windows on (V$_{1-x}$Cr$_x$)$_2$O$_3$, $x=0$ and (110)-oriented surface (a) at $\lambda$ = 800 nm (b) at $\lambda$ = 800 and 1550 nm (the reflectivity scale, vertical, is cut in order to expand the coherent acoustic oscillation).} 
\label{complete}
\end{center}
\end{figure}

In Fig.~\ref{complete} we present a typical example of a time-resolved reflectivity curve on (V$_{1-x}$Cr$_x$)$_2$O$_3$ at $\lambda$=800 nm, taken on an undoped sample, oriented so that the (110) direction in real-space hexagonal notation was perpendicular to the surface (and hence almost parallel to the laser light wavevector). The transient response can be decomposed in three parts. 

First, at zero time delay, so just after the arrival of the excitation pump on the system, we observe an ultrashort negative peak, lasting about 100 fs. This ultrafast drop of reflectivity (the minimum value is labelled $R_1$ in Fig.~\ref{complete} (a)) corresponds to the excitation and relaxation of electrons, and we found that its amplitude is about twice as large in the metallic phase with respect to the insulating one, since the presence of a gap in the Mott insulator reduces the number of photoexcited electrons. This peak ended at a reflectivity value labeled $R_2$ in Fig.~\ref{complete} (a) and (b).

It is followed by two coherent oscillations. The first one, shown in Fig.~\ref{complete} (a), is due to the presence of a coherent optical phonon \cite{Misochko1998}. $R_2$ somehow establishes the frontier between the sub-ps and ps temporal regions in our curves, and the corresponding time delay is related to the electron-phonon thermalisation time in V$_2$O$_3$. The reasons for determining its value will be further discussed later.

At longer time scales (Fig.~\ref{complete} (b)) another oscillation takes place, which we attributed to the propagation of a coherent acoustic wave \cite{Mansart2010}, since its frequency corresponds to a wave with the same wavelength as the laser pulse but travelling at the velocity of sound in V$_2$O$_3$ \cite{Yelon1979, Seikh2006}. To further corroborate this interpretation, we studied the same specimen at a wavelength $\lambda$ = 1550 nm, finding an approximately double wave period and thus confirming the coherent acoustic nature of this oscillation.  

The excitation of coherent acoustic waves by an ultrashort laser pulse was phenomenologically described by Thomsen \cite{Thomsen1986}. The detection of coherent acoustic waves in a pump-probe reflectivity measurement can be explained as a Brillouin scattering \cite{Brillouin1922} phenomenon occurring in the system after excitation of the wave by the pump pulse. The scattering condition is 

\begin{equation}
\label{brillscatt}
q_{phonon}= 2 n k_{probe}cos\left(\theta_i\right)
\end{equation}

where $q_{phonon}$ is the phonon wavevector, $n$ the real part of the refractive index, and the probe photon has a wavevector $k_{probe}$ arriving at an incidence angle $\theta_{i}$ with respect to the surface normal. Following this scattering condition, the probe beam acts as a filter to select the acoustic wave propagating along the scattering plane symmetry axis, i.e. the normal to the sample surface, and travelling with the wavevector $q_{phonon}$. 

In this work, we explore in detail two issues which have not been investigated so far and which are crucially important for a prototype Mott-Hubbard material, namely the effects of the thermodynamic phase and more specifically of the correlated part of the electron density of states on the picosecond time scale transient reflectivity. This was possible thanks to a detailed study of the coherent acoustic oscillation while spanning the temperature-doping phase diagram of (V$_{1-x}$Cr$_x$)$_2$O$_3$ across the metal-insulator transition. 

As previously reported~\cite{Mansart2010}, the experimental geometry can strongly affect the coherent acoustic response in complex materials: consequently, in this study we paid particular attention to compare specimens which had all the same crystallographic orientation with respect to the laser beams. In particular, for all the samples the c axis was in the plane of the sample surface and parallel to the electric field of the probe laser beam.

In Fig.~\ref{diff_pmpi}, we present the pump-probe reflectivity in the 0-40 ps time window for the insulating ($x=0.028$) and the metallic phase ($x=0$) and for four different laser fluences, at $\lambda$ = 800 nm. Changing the laser wavelength to 1550 nm did not qualitatively change our results, indicating that the electronic excitation are d-d transitions between vanadium 3d levels, reachable both at 0.8 and 1.55 eV. The comparison between Fig.~\ref{diff_pmpi} (a) and (b) clearly indicates that the response is quite different between the two samples, but a more careful analysis can show that the difference is mostly due to a constant offset in reflectivity, while the coherent acoustic oscillation is very similar in the two cases. 

In order to discriminate between coherent lattice oscillations and electronic effects, we can start observing the fluence dependence of the curves of the $x=0.028$ sample. The coherent acoustic oscillations are particularly evident thanks to the presence of three nodal regions at about 3, 12 and 28 ps. Since with a reasonable approximation one can connect these three nodal points with a line parallel to the abscissa axis, we fitted our curves with the following function ~\cite{Mansart2010}:  

\begin{equation}
\label{acousticwave}
\frac{\Delta R}{R}=A_{ac}sin\left(\frac{4\pi n v_{s}cos\theta_{i}}{\lambda_{probe}}\left(1+ht\right)t+\phi\right)e^{-t/\beta}+A_{0}
\end{equation}

where $n$ is the real part of the refraction index, $\lambda_{probe}$ the laser wavelength and $\theta_i$ the probe beam incidence angle; the detected acoustic wave propagates at the sound velocity $v_s$, with a wavevector corresponding to the Brillouin scattering condition. It has an amplitude $A_{ac}$, a phase $\phi$ and a lifetime $\beta$. $A_{0}$ represents the mean value of its oscillation, that we assume here to be a constant baseline. Since the time difference between the second and third nodal point is larger than between the first and the second one, we assumed that the frequency depends on the time delay (even though some caution is needed due to the very small number of oscillations visible before the damping of the acoustic wave). Thus, we used as fitting frequency the function $f_0(1+ht)$, imposing a 
common $f_0$ parameter for all pump fluences and a given sample, while $h$ was allowed to slightly vary with fluence around a value of $-6.5\times10^{-3}$. We obtained a frequency value close to 52 GHz at $\lambda$ = 800 nm and 21 GHz at $\lambda$ = 1550 nm, compatible with previous measurements of sound velocity ($7.47\times10^5~cm s^{-1}$ in sample oriented in (100)-reciprocal direction~\cite{Yelon1979}) in (V$_{1-x}$Cr$_x$)$_2$O$_3$ for the orientations we studied.

	\begin{widetext}

\begin{center} 
\includegraphics[angle=0,width=0.75\linewidth,clip=true]{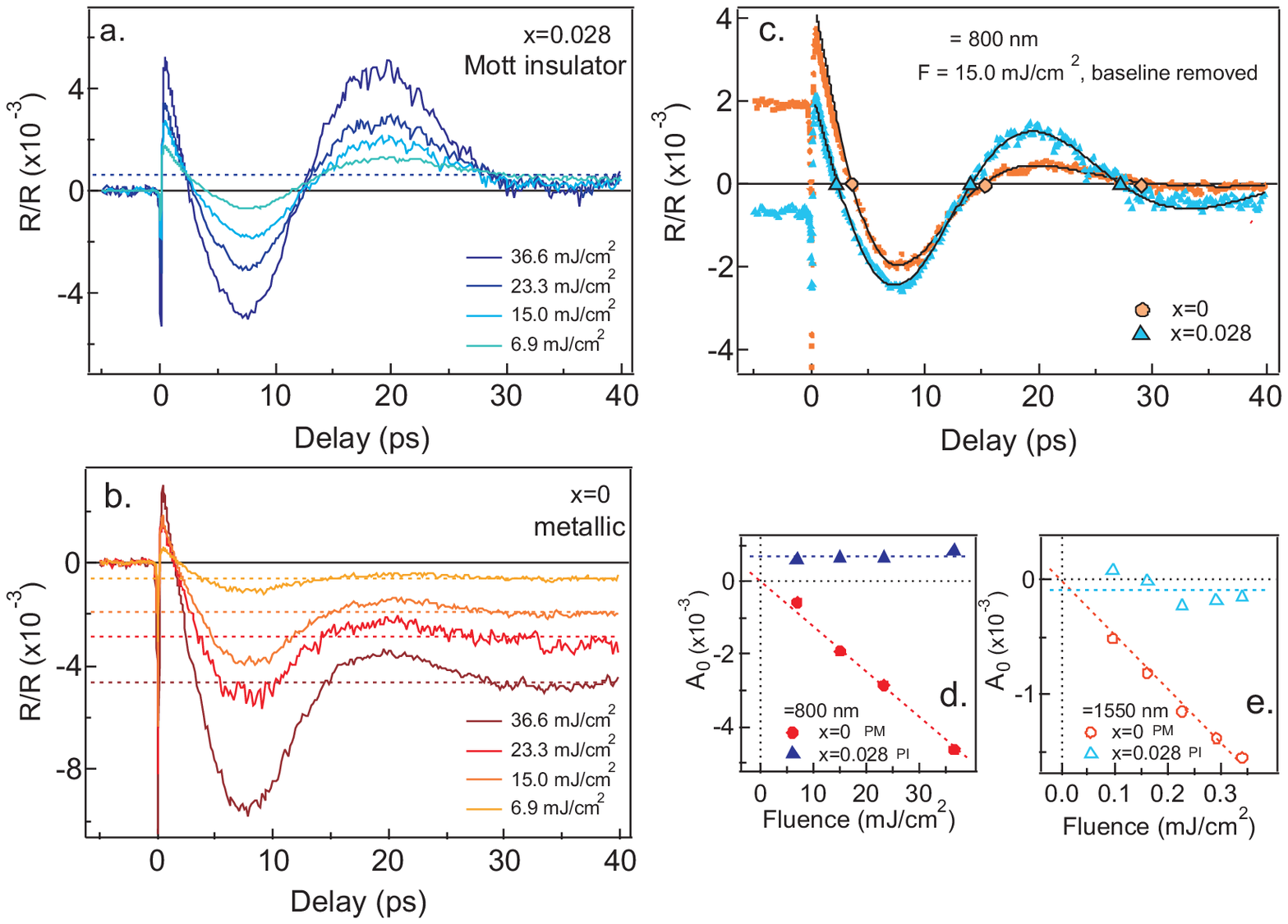} 
\caption{(Colour online) Thermodynamic phase effect on (V$_{1-x}$Cr$_x$)$_2$O$_3$ transient reflectivity, at $\lambda$ = 800 nm. Oscillation amplitudes increase with laser fluence, and the mean values $A_{0}$ are represented with dotted lines in (a) and (b).(c) Transient reflectivity curves fitting procedure (the oscillation baseline has been removed in order to compare the coherent oscillation in the two phases) (d) Oscillation mean value A$_0$ as a function of the pump fluence for $\lambda$ = 800 nm, and (e) for $\lambda$ = 1550 nm. Dashed lines in (d) and (e) are linear fits of the data points.}
\label{diff_pmpi} 
\end{center}

	\end{widetext} 
	
We found that this fitting function could be well used also for the metallic sample, if one allows the mean value of the oscillation $A_0$ to vary with the laser fluence. 
The actual similarity between the two cases ($x=0$ and $x=0.028$) is well visible in Fig.~\ref{diff_pmpi} (c), where we removed the oscillation baseline $A_0$ in order to show solely the coherent acoustic wave, and to better compare it in the two thermodynamic phases for the same laser fluence: the two waves present a remarkably similar behaviour.

The sinusoidal fitting curves are also shown in Fig.~\ref{diff_pmpi} (c): to obtain them, we first fitted each spectrum only between the first and the third node, which are the intersections of the experimental curve and the zero line after removing the oscillation baseline. With the fitting parameters determined in this way, we constructed the solid lines in Fig.\ref{diff_pmpi} (c) over the whole time window, finding that the agreement with the experimental points is very good everywhere. This is particularly noteworthy for the region between t=0 and the first node, because this means that the value of the reflectivity in this region (i.e. the value $R_2$ as indicated in Fig.~\ref{complete}) is essentially determined by the acoustic oscillation amplitude. 

As a function of the thermodynamic phase, there are some slight differences on the coherent waves themselves. The acoustic phonon amplitude $A_{ac}$ is roughly identical in the metal and in the insulator and its value is proportional to the pump fluence. Its lifetime $\beta$ is longer in the insulator than in the metal, for all pump fluences and wavelengths: this is not surprising, since in the metallic phase the presence of free electrons tends to accelerate the damping of coherent lattice oscillations. $\beta$ increases while reducing the pump fluence, from 19 ps to 40 ps ($x=0.028$) or from 9 ps to 30 ps ($x=0$). This increase reflects the diminution of phonon-phonon scattering, which induce a damping of the coherent oscillation, while reducing the number of coherent phonons.  

If the coherent acoustic oscillations are very similar, the main difference between the two phases on the ps time scale are the baselines of these oscillations, as presented in Fig.~\ref{diff_pmpi} (d) and (e). For the Cr-2.8\% doped sample, for all pump fluences and wavelengths, the oscillation takes place roughly around the same mean value $A_{0}$, very close to zero,  corresponding to the dotted baseline shown in Fig.~\ref{diff_pmpi} (a).
On the other hand, the metallic phase displays different mean values: the curves are shifted towards negative reflectivity values, and this shift - which 
corresponds to the value labelled $R_3$ in the curve of Fig.~\ref{complete} (b) - increases with the pump fluence. 

The fact that in all our data the oscillation mean value can be well approximated by a constant $A_{0}$ up to at least 130 ps implies that the typical time for the reflectivity to recover its ground state value must be at least ot the order of nanoseconds, corresponding to thermal relaxation of the lattice.

The comparison between the two specimens in Fig.~\ref{diff_pmpi} (d) and (e) clearly shows that in the metallic phase the lattice is more efficiently heated by the laser excitation, and that the evolution of $A_{0}$ is consistent with a linear behaviour vs. fluence. We emphasize that our conclusion on the difference between metal and insulator is valid independently of the fits in Fig.~\ref{diff_pmpi} (c), which are the result of a very simplified vision of the problem: in particular, in our case a time-independent mean value was adopted to fit the coherent acoustic oscillations. But even if the mean value $A_{0}$ depended on the time delay, we believe that the conclusion we extract here on the interplay between coherent oscillations and electronic effects would still have a general interest, beyond the specific case of vanadium sesquioxide: the dependence on fluence of the oscillation baseline changes between the metal and the insulator, and appears to be a signature of the thermodynamic phase in ultrafast pump-probe experiments. 

\begin{figure}[h!] 
\begin{center} 
\includegraphics[angle=0,width=1\linewidth,clip=true]{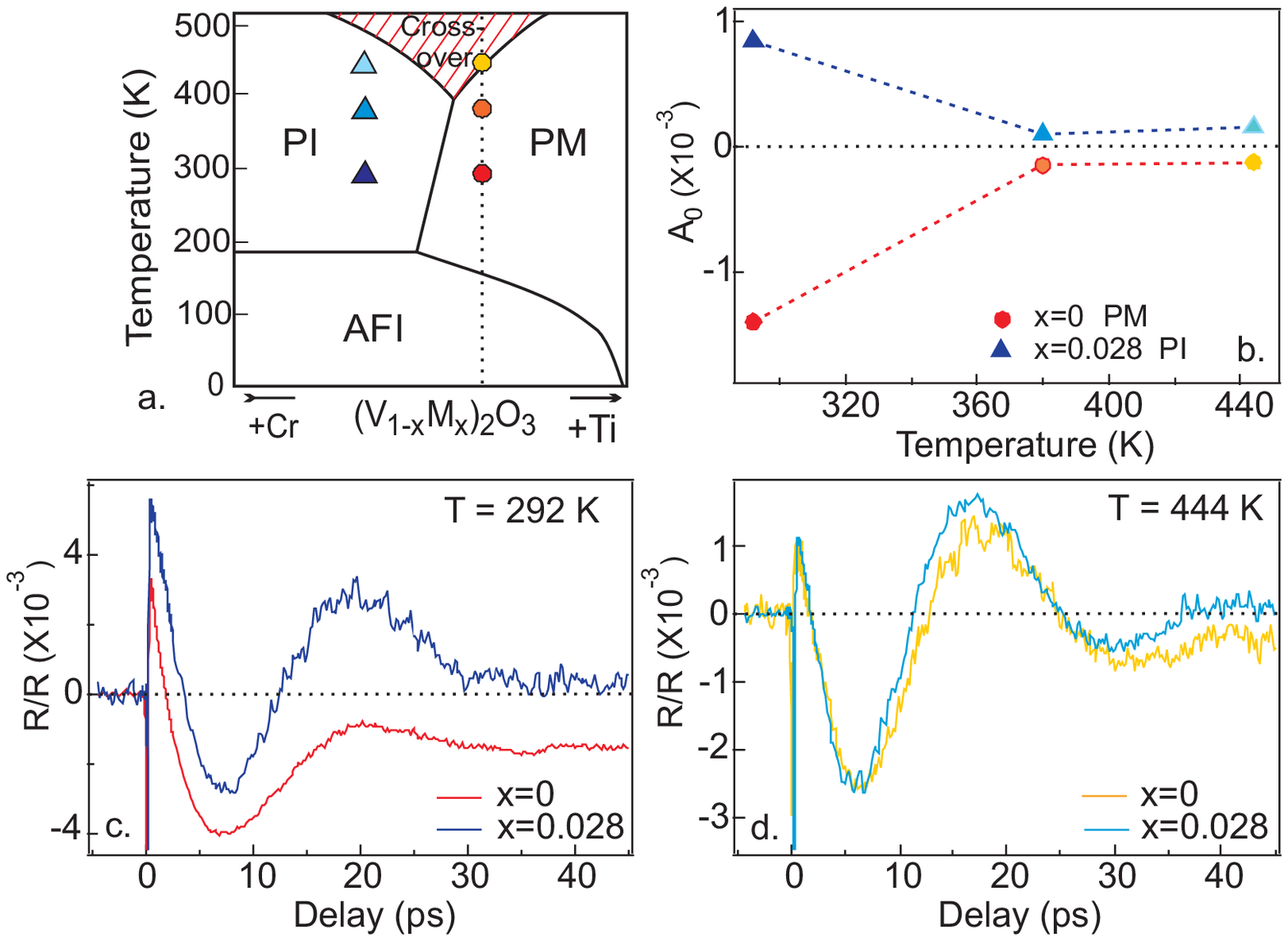} 
\caption{(Colour online) (a) Phase diagram of (V$_{1-x}$M$_x$)$_2$O$_3$, M = Ti, Cr. The dashed line represents the undoped compound ($x=0$), and the Mott transition occurs between the PI and the PM phases; the dashed region is the cross-over of the Mott transition. (b) Behaviour of the oscillation mean value $A_{0}$ versus temperature for PM (circles) and PI (triangles). (c) Transient reflectivity for $x=0$ and $x=0.028$, at T=292 K and (d) at T=444 K.} 
\label{chaud} 
\end{center} 
\end{figure} 

The question that naturally arises from this first conclusion is to what extent it is determined by the fact that the metal-insulator transition discussed here is a Mott transition, i.e. determined by electron correlation effects. The most natural way to answer this question is to move across the phase diagram by changing the temperature: in fact, in a Mott-Hubbard system temperature markedly affects the electron density of states, in particular for the correlated metal: as recently demonstrated for  V$_2$O$_3$ with optical~\cite{Baldassarre2008} and photoelectron spectroscopy ~\cite{Rodolakis2009}, the coherent quasiparticle weight in the PM phase is strongly  reduced as the temperature is increased to 400-450K. At these temperatures, the system approaches and enters the cross-over region, where the distinction between metallic and insulating states vanishes, as recently investigated by means of transport~\cite{Limelette2003} experiments. We therefore performed transient reflectivity measurements at different temperatures, and the comparison between PM and PI is presented for T=292 K (Fig.~\ref{chaud} (b)) and T=444 K (Fig.~\ref{chaud} (d)). 

Very interestingly, the differences observed between PM and PI at room temperature, well below the cross-over region (difference in oscillation lifetime and baseline), almost disappear while increasing the temperature: as shown in Fig.~\ref{chaud} (d), very close to the cross-over part of the phase diagram, the two specimens show a similar response. This unambiguously indicates how the behaviours of the mean value and of the lifetime of the coherent acoustic wave are determined by the thermodynamic phase. In particular, the values of the oscillation mean value $A_{0}$ versus  temperature are presented in the Fig.~\ref{chaud} (b): one can observe how the change is particularly evident for the PM phase, suggesting that this should be  related to the suppression of the coherent quasiparticle spectral weight observed with other spectroscopies~\cite{Baldassarre2008,Rodolakis2009}. Therefore, one can conclude that the ultrafast spectroscopic signatures we discussed so far  are not only determined by the thermodynamic phase (PM/PI), but are intrinsically related to the correlated electron density of states - in particular to its coherent part, which strongly changes with temperature within the metallic phase. 

One additional but important conclusion which can be drawn from this discussion is that all these effects, evolving on a ps time scale and interesting on their own, must be taken into account to explain the response of the systems even in the sub-ps time window. The electronic excitation peak and its following relaxation are studied in detail in all pump-probe experiments, the former being related to the number of photoexcited electrons, the latter to the electron-phonon coupling constant determining electron-lattice thermalisation \cite{Allen1987}. As we have discussed, the value of $R_2$ (Fig.~\ref{complete}) is affected by many parameters (thermodynamic phase through a rigid shift of the reflectivity, acoustic wave amplitude depending on cristallographic orientation~\cite{Mansart2010}), and this of course affects the shape of the electronic excitation peak, in particular of its relaxation (the transition from $R_1$ to $R_2$): this may be not clear at all if just the peak itself is studied, within less than 1 ps from photoexcitation. For instance, according to the analysis performed, one might decide to remove the coherent acoustic oscillation amplitude from the mesured signal, to keep only its mean value (which in the case of V$_2$O$_3$ studied here corresponds to $R_3$), but in order to do that correctly it is necessary to have a general vision of the ultrafast response over tenths of ps. We also point out that this can be of relevance for other strongly correlated materials where for instance the metal-insulator transition implies symmetry breaking, such as Peierls transitions where the lattice symmetry is lowered, or N\'{e}el or Curie transitions where magnetic order takes place, resulting on the breaking of the time-reversal symmetry. In particular, care should be taken not to mistake the effects presented here (coherent acoustic oscillation, rigid shift due to the metal-insulator transition) for signatures of other phenomena occurring in the material. 

\section{Conclusion}
\label{concl}

In summary, we studied the ultrafast response of the prototype Mott compound (V$_{1-x}$Cr$_x$)$_2$O$_3$ using fs time resolved pump-probe reflectivity. 
Two different Cr doping levels, corresponding to the Paramagnetic Metallic ($x=0$) and Paramagnetic Insulating ($x=0.028$) phases, were extensively investigated for various temperatures, laser fluences and wavelengths. This allowed us to disentangle the effects of the thermodynamic phase and of coherent lattice oscillations in the overall ultrafast reflectivity signal. In particular, the mean value of the transient reflectivity gives a clear spectroscopic signature of the thermodynamic phase on the ps time scale, as shown by a detailed study as a function of the position in the phase diagram. The correlated metal shows the most pronounced changes while approaching the cross-over regime of the systems, suggesting that these changes are intimately related to the correlated electron density of states, in particular to the coherent quasiparticle spectral weight. 
On the sub-ps time scale, we showed how these effects, in particular the coherent acoustic oscillations, can affect the genuine lineshape of the electronic excitation peak, which is normally studied to extract important physical parameters like the electron-phonon coupling: this clearly shows that the study of the transient regime of the material over several ps is critically important to correctly interpret its time evolution also in the sub-ps time window.  
These results, obtained by studying a model isostructural metal-insulator transition, can be of relevance to interpret the interplay between coherent lattice oscillations and thermodynamic phase effects also for other strongly correlated materials presenting more complex phase transitions.

\acknowledgments
The authors would like to thank Patricia Metcalf for providing high quality vanadium sesquioxide single crystals, and Pawel Wzietek for interesting and  stimulating discussions. The expert help of Gilles Rey and the continuous and enthousiastic support of Pierre-Antoine Albouy are gratefully acknowledged.  

\bibliographystyle{unsrt}

\end{document}